\documentstyle[aps,graphicx]{revtex}\tighten
\draft
\def \D {\mbox{D}}

\begin{document}
\twocolumn[\hsize\textwidth\columnwidth\hsize\csname
@twocolumnfalse\endcsname

\title{Density perturbations in the brane-world}

\author{Christopher Gordon
 { }and Roy Maartens
}
\address{~}
\address{Relativity and Cosmology Group, School of
Computer Science and Mathematics, Portsmouth University,
Portsmouth~PO1~2EG, Britain}

\maketitle

\begin{abstract}

In Randall-Sundrum-type brane-world cosmologies, density
perturbations generate Weyl curvature in the bulk, which in turn
backreacts on the brane via stress-energy perturbations. On large
scales, the perturbation equations contain a closed system on the
brane, which may be solved without solving for the bulk
perturbations. Bulk effects produce a non-adiabatic mode, even
when the matter perturbations are adiabatic, and alter the
background dynamics. As a consequence, the standard evolution of
large-scale fluctuations in general relativity is modified. The
metric perturbation on large-scales is {\em not} constant during
high-energy inflation. It is constant during the radiation era,
except at most during the very beginning, if the energy is high
enough.

\end{abstract}

\pacs{04.50.+h, 98.80.Cq \hfill{hep-th/0009010}}
\vskip2pc]

\section{Introduction}

According to string and M-theory, gravity is a higher-dimensional
theory, reducing to Einstein's four-dimensional theory of general
relativity at low enough energies. In the brane-world scenario,
the standard model matter fields are confined to a 3-brane in
$1+3+d$ dimensions, while the gravitational field can propagate in
the bulk, i.e., also in the $d$ extra dimensions, being localized
at the brane at low energies. Recent developments show that the
$d$ extra space dimensions need not be small, or even compact,
thus allowing the intriguing possibility that corrections could
occur even at TeV scales.

These exciting theoretical developments may offer a promising
route towards a quantum gravity theory. However, as well as
theoretical elegance, they must also pass the increasingly
stringent tests provided by cosmological observations. Primarily,
this involves developing higher-dimensional perturbation theory
and then applying it to analyze the generation and evolution of
density and tensor perturbations on the brane, leading to a
prediction of the CMB (cosmic microwave background) anisotropies
and galaxy distribution.

This is an ambitious and difficult program, but initial steps have
already been taken, at least in the case of a particular class of
models that generalize the Randall-Sundrum models~\cite{rs}.
Large-scale adiabatic density perturbations from inflation on the
brane have been computed~\cite{mwbh} (see also~\cite{cll}), using
the conservation of the curvature perturbation on uniform-density
hypersurfaces. This conservation follows from adiabaticity and the
conservation of energy-momentum on the brane, and is independent
of the form of the field equations~\cite{wmll}. In~\cite{mwbh},
the backreaction effect of metric fluctuations in the fifth
dimension was neglected. In the general case, i.e., incorporating
also the fluctuations in the nonlocal quantities that carry the
bulk influence onto the brane, it has been shown that {\em
large-scale scalar perturbations contain a closed system on the
brane}---and thus can in principle be evaluated purely from
initial conditions on the brane, without knowledge of bulk
dynamics~\cite{m}. In this paper, we solve the closed system given
in~\cite{m} to find the evolution of large-scale density
perturbations on the brane.

A general perturbation formalism has been
developed~\cite{pert,bd}, encompassing equations on the brane and
in the bulk, and in principle able to describe all scales.
However, the general equations are extremely complicated. A first
application of the equations has been made to large-scale tensor
perturbations from inflation on the brane~\cite{lmw}. Unlike the
scalar case, large-scale tensor perturbations cannot be evaluated
without the bulk perturbation equations.

We develop the outline argument presented first in~\cite{m}, and
analyze large-scale density perturbations and their evolution, from
after Hubble-crossing in inflation through the radiation era. This
provides the basis for predicting the large-angle scalar
anisotropies generated in CMB temperature, and seeing how the bulk
effects modify general relativistic predictions. We show that in
general, the perturbation $\Phi$ (a covariant analog of the
Bardeen metric perturbation) is {\em no longer constant during
high-energy inflation, but grows.} However, $\Phi$ is constant during the
radiation era, as in general relativity, except at most in the
early radiation era, if the energy density is still high relative
to the brane tension. This means that {\em the Sachs-Wolfe plateau
in the CMB anisotropies is likely to be preserved,\footnote{
The Sachs-Wolfe formula has not yet been calculated in brane-world
cosmologies, but we expect that the corrections to the general
relativistic result will be very small, since the energy scale at
last scattering is much less than the brane tension.
} but the limits which COBE measurements place on inflationary
potentials will be changed, and could also become sensitive to the
form of the potential.}

\section{Brane dynamics}

We follow the approach and notation of~\cite{m}. The 5-dimensional
(bulk) field equations are
\begin{equation}
\widetilde{G}_{AB} =
\widetilde{\kappa}^2\left[-\widetilde{\Lambda}\widetilde{g}_{AB}
+\delta(\chi)\left\{ -\lambda g_{AB}+T_{AB}\right\}\right]\,,
\label{1}
\end{equation}
where tildes denote the bulk generalization of standard general
relativity quantities, and $\widetilde{\kappa}^2=
8\pi/\widetilde{M}_{\rm p}^3$, where $\widetilde{M}_{\rm p}$ is
the fundamental 5-dimensional Planck mass, which is typically much
less than the effective Planck mass on the brane, $M_{\rm
p}=1.2\times 10^{19}$ GeV. The brane is given by $\chi=0$, so that
a natural choice of coordinates is $x^A=(x^\mu,\chi)$, where
$x^\mu=(t,x^i)$ are spacetime coordinates on the brane. The brane
tension is $\lambda$, and $g_{AB}=\widetilde{g}_{AB}-n_An_B$ is
the induced metric on the brane, with $n_A$ the spacelike unit
normal to the brane. Standard-model matter fields confined to the
brane make up the brane energy-momentum tensor $T_{AB}$ (with
$T_{AB}n^B=0$). The bulk cosmological constant
$\widetilde{\Lambda}$ is negative, and is the only 5-dimensional
stress energy. (See~\cite{mw} for the modification of this
approach in the case where there is also a scalar field in the
bulk.)

The most general background bulk with homogeneous and isotropic
induced metric on the brane is the 5-dimensional
Schwarzschild-anti-de Sitter metric~\cite{msm}, with the black
hole mass leading to an effective radiation-like correction to the
Friedmann equation on the brane~\cite{bdel}.

The field equations induced on the brane are derived via an
elegant geometric approach in~\cite{sms}, leading to new terms
that carry bulk effects onto the brane:
\begin{equation}
G_{\mu\nu}=-\Lambda g_{\mu\nu}+\kappa^2
T_{\mu\nu}+\widetilde{\kappa}^4S_{\mu\nu} - {\cal E}_{\mu\nu}\,,
\label{2}
\end{equation}
where $\kappa^2=8\pi/M_{\rm p}^2$. The various energy scales are
related to each other via
\begin{equation}
\lambda=6{\kappa^2\over\widetilde\kappa^4} \,, ~~ \Lambda =
{\textstyle{1\over2}}\widetilde\kappa^2\left(\widetilde{\Lambda}+
{\textstyle{1\over6}}\widetilde\kappa^2\lambda^2\right)\,,
\label{3}
\end{equation}
and the high-energy regime is $\rho \gtrsim \lambda$. Bulk
corrections to the Einstein equations on the brane are of two
forms: firstly, {\em the matter fields contribute local quadratic
energy-momentum corrections} via the tensor $S_{\mu\nu}$, and
secondly, there are {\em nonlocal effects from the free
gravitational field in the bulk,} transmitted via the projection
${\cal E}_{\mu\nu}$ of the bulk Weyl tensor. The matter
corrections are given by
\begin{eqnarray}
S_{\mu\nu}&=&{\textstyle{1\over12}}T_\alpha{}^\alpha T_{\mu\nu}
-{\textstyle{1\over4}}T_{\mu\alpha}T^\alpha{}_\nu\nonumber\\
&&~~{}+ {\textstyle{1\over24}}g_{\mu\nu} \left[3 T_{\alpha\beta}
T^{\alpha\beta}-\left(T_\alpha{}^\alpha\right)^2 \right]\,.
\label{3'}
\end{eqnarray}
The projection of the bulk Weyl tensor is
\begin{equation}
{\cal E}_{AB}=\widetilde{C}_{ACBD}n^C n^D\,,
\label{4}
\end{equation}
which is symmetric and traceless and without components orthogonal
to the brane, so that ${\cal E}_{AB}n^B=0$ and ${\cal E}_{AB}\to
{\cal E}_{\mu\nu}g_A{}^\mu g_B{}^\nu$ as $\chi\to 0$.

The Weyl tensor $\widetilde{C}_{ABCD}$ represents the free,
nonlocal gravitational field in the bulk, i.e., the part of the
field that is not directly determined at each point by the
energy-momentum tensor at that point. The local part of the bulk
gravitational field is the Einstein tensor $\widetilde{G}_{AB}$,
which is determined locally via the bulk field equations
(\ref{1}). Thus ${\cal E}_{\mu\nu}$ transmits nonlocal
gravitational degrees of freedom from the bulk to the brane,
including tidal (or Coulomb), gravito-magnetic and transverse
traceless (gravitational wave) effects~\cite{m}.

If $u^\mu$ is the 4-velocity comoving with matter (which we assume
is a perfect fluid or minimally-coupled scalar field), we can
decompose the nonlocal term as
\begin{equation}
{\cal E}_{\mu\nu}={-6\over\kappa^2\lambda}\left[{\cal
U}\left(u_\mu u_\nu+{\textstyle {1\over3}} h_{\mu\nu}\right)+{\cal
P}_{\mu\nu}+{\cal Q}_{\mu}u_{\nu}+{\cal Q}_{\nu}u_{\mu}\right]\,,
\label{6}
\end{equation}
where $h_{\mu\nu}=g_{\mu\nu}+u_\mu u_\nu$ projects into the
comoving rest-space. Here
\[
{\cal U}=-{\textstyle{1\over6}}\kappa^2 \lambda\, {\cal
E}_{\mu\nu}u^\mu u^\nu
\]
is {\em an effective nonlocal energy density} on the brane (which
need not be positive), arising from the free gravitational field
in the bulk. It carries Coulomb-type effects from the bulk onto
the brane. There is {\em an effective nonlocal anisotropic stress}
\[
{\cal P}_{\mu\nu}=-{\textstyle{1\over6}}\kappa^2 \lambda\left[
h_\mu{}^\alpha h_\nu{}^\beta-{\textstyle{1\over3}}h^{\alpha\beta}
h_{\mu\nu}\right] {\cal E}_{\alpha\beta}
\]
on the brane, which carries Coulomb, gravito-magnetic and
gravitational wave effects of the free gravitational field in the
bulk. The {\em effective nonlocal energy flux} on the brane,
\[
{\cal Q}_\mu ={\textstyle{1\over6}}\kappa^2 \lambda\,
h_\mu{}^\alpha {\cal E}_{\alpha\beta}u^\beta\,,
\]
carries Coulomb and gravito-magnetic effects from the free
gravitational field in the bulk. (Note that there is no energy
flux in the bulk, and thus no transfer of energy between bulk and
brane; this situation changes if bulk scalar fields are
present~\cite{bd,mw}.)

\section{Local and nonlocal conservation equations}

The local and nonlocal bulk modifications may be consolidated into
an effective total energy-momentum tensor:
\begin{equation}
G_{\mu\nu}=-\Lambda g_{\mu\nu}+\kappa^2 T^{\rm tot}_{\mu\nu}\,,
\label{6'}
\end{equation}
where
\begin{equation}
T^{\rm tot}_{\mu\nu}= T_{\mu\nu}+{6\over \lambda}S_{\mu\nu}-
{1\over\kappa^2}{\cal E}_{\mu\nu}\,. \label{6''}
\end{equation}
The effective total energy density, pressure, anisotropic stress
and energy flux are
\begin{eqnarray}
\rho^{\rm tot} &=& \rho\left(1+{\rho\over2\lambda}\right)+{6 {\cal
U}\over\kappa^4\lambda}\,, \label{a}\\ p^{\rm tot} &=& p+
{\rho\over2\lambda}(\rho+2p) +{2{\cal U}\over\kappa^4\lambda}\,,
\label{b}\\ \pi^{\rm tot}_{\mu\nu} &=&{6\over
\kappa^4\lambda}{\cal P}_{\mu\nu}\,, \label{c}\\ q^{\rm tot}_\mu
&=& {6\over \kappa^4\lambda}{\cal Q}_\mu \,.\label{d}
\end{eqnarray}

The brane energy-momentum tensor separately satisfies the
conservation equations, $\nabla^\nu T_{\mu\nu}=0 $, giving
\begin{eqnarray}
&&\dot{\rho}+\Theta(\rho+p)=0\,,\label{pc1}\\ && \D_\mu
p+(\rho+p)A_\mu =0\,,\label{pc2}
\end{eqnarray}
where a dot denotes $u^\nu\nabla_\nu$, $\Theta=\D^\mu u_\mu$ is
the volume expansion rate of the $u^\mu$ congruence,
$A_\mu=\dot{u}_\mu$ is its 4-acceleration, and $\D_\mu$ is the
projected covariant spatial derivative. The Bianchi identities on
the brane imply that the projected Weyl tensor obeys the
constraint
\begin{equation}
\nabla^\mu{\cal E}_{\mu\nu}={6\kappa^2\over\lambda}\nabla^\mu
S_{\mu\nu}\,. \label{5}
\end{equation}
This shows how nonlocal bulk effects are sourced by local bulk
effects, which include spatial gradients and time derivatives:
{\em evolution and inhomogeneity in the matter fields can generate
nonlocal gravitational effects in the bulk, which backreact on the
brane.} The brane energy-momentum tensor and the consolidated
effective energy-momentum tensor are {\em both} conserved
separately. Conservation of $T^{\rm tot}_{\mu\nu}$ gives, upon
using Eqs. (\ref{a})--(\ref{pc2}), propagation equations for the
nonlocal energy density ${\cal U}$ and energy flux ${\cal Q}_\mu$.
In linearized form, these are
\begin{eqnarray}
&& \dot{\cal U}+{\textstyle{4\over3}}\Theta{\cal U}+\D^\mu{\cal
Q}_\mu  =0 \,, \label{lc1'}\\&& \dot{\cal Q}_{\mu}+4H{\cal Q}_\mu
+{\textstyle{1\over3}}\D_\mu{\cal U}+{\textstyle{4\over3}}{\cal
U}A_\mu \nonumber\\&&~~~{}+\D^\nu{\cal P}_{\mu\nu}
=-{\textstyle{1\over6}} \kappa^4(\rho+p) \D_\mu \rho
\,,\label{lc2'}
\end{eqnarray}
where $H=\dot{a}/a$ ($={1\over3}\Theta$) is the Hubble rate in the
background. The nonlocal tensor mode, which satisfies $\D^\nu{\cal
P}_{\mu\nu}=0 \neq {\cal P}_{\mu\nu}$, does not enter the nonlocal
conservation equations. Furthermore, there is no evolution
equation at all for ${\cal P}_{\mu\nu}$, reflecting the fact that in
general the equations do not close on the brane, and one needs
bulk equations to determine brane dynamics. There are bulk degrees
of freedom whose impact on the brane cannot be predicted by brane
observers. {\em The evolution of the nonlocal energy density and
flux, which carry scalar and vector modes of the bulk
gravitational field, is determined on the brane, while the
evolution of the nonlocal anisotropic stress, which carries
scalar, vector and tensor modes of the bulk field, is not.}

The generalized Raychaudhuri equation on the brane in linearized
form is
\begin{eqnarray}
&&\dot{\Theta}+{\textstyle{1\over3}}\Theta^2 -{\rm D}^\mu
A_\mu+{\textstyle{1\over2}}\kappa^2(\rho + 3p) -\Lambda
\nonumber\\&&~~{}= -{\textstyle{1\over2}} \kappa^2
(2\rho+3p){\rho\over\lambda} -{6 {\cal U}\over\kappa^2\lambda}\,,
\label{prl}
\end{eqnarray}
where the general relativistic case is recovered when the
right-hand side is set to zero. In the background, this gives
\begin{eqnarray}
\dot{H}&=&-H^2-{\kappa^2\over6}\left[\rho+3p+{\rho\over 2\lambda}
(2\rho+3p)\right]\nonumber\\&&~~{}+{1\over3}\Lambda- {2\over
\kappa^2\lambda }{\cal U}_o\left({a_o\over
a}\right)^4\,,\label{bray}
\end{eqnarray}
where the solution for ${\cal U}$ follows from Eq.~(\ref{lc1'}),
$a_o$ is the initial scale factor and ${\cal U}_o={\cal U}(a_o)$.
The first integral of this equation is the generalized Friedmann
equation on the brane:
\begin{equation}\label{f}
H^2={\kappa^2\over3} \rho\left(1 +{\rho\over2\lambda}\right) +
{1\over3}\Lambda -{K\over a^2}+ {2\over \kappa^2\lambda } {\cal
U}_o\left({a_o\over a}\right)^4\,,
\end{equation}
where $K=0,\pm1$. Local bulk effects modify the background
dynamics. In particular, inflation at high energies
($\rho\gtrsim\lambda$) proceeds at a higher rate than the
corresponding rate in general relativity. This introduces
important changes to the dynamics of the early
universe~\cite{mwbh,cll,eu}, and accounts for an increase in the
amplitude of scalar and tensor fluctuations at
Hubble-crossing~\cite{mwbh,lmw}.

The condition for inflation becomes~\cite{mwbh}
\begin{equation}\label{inf}
w<-{1\over3}\left({2\rho+\lambda\over \rho+\lambda} \right)\,,
\end{equation}
where $w=p/\rho$. As $\rho/\lambda\to\infty$, we have
$w<-{2\over3}$, while the general relativity condition
$w<-{1\over3}$ is recovered as $\rho/\lambda\to 0$.

If ${\cal U}_o=0$, i.e., if the background bulk is conformally
flat, then Eqs.~(\ref{a}) and (\ref{b}) show that the effective
equation of state index for the total energy-momentum tensor is
\begin{equation}\label{w}
w^{\rm tot}\equiv {p^{\rm tot}\over \rho^{\rm tot}}
={w+(1+2w)\rho/2\lambda \over 1+\rho/2\lambda}\approx 1+2w\,,
\end{equation}
where the last equality holds at very high energies
($\rho\gg\lambda$). Thus for slow-roll inflation, $w^{\rm tot}$
and $w$ are both close to $-1$. The high-energy inflation
condition $w<-{2\over3}$ is $w^{\rm tot}<-{1\over3}$. During
high-energy reheating with $w\approx 0$ on average, we have
$w^{\rm tot}\approx 1$, so that the effective equation of state is
stiff, while high-energy radiation-domination ($w={1\over3}$) has
$w^{\rm tot}\approx{5\over3}$, i.e., an ultra-stiff effective
equation of state. The effective sound speed at very high energies
is also altered:
\begin{equation}\label{cs}
(c_{\rm s}^2)^{\rm tot}\equiv {\dot{p}^{\rm tot} \over
\dot{\rho}^{\rm tot}} \approx c_{\rm s}^2+w+1\,,
\end{equation}
where $c_{\rm s}^2=\dot{p}/\dot{\rho}$.

\section{Scalar perturbations on the brane}

Scalar perturbations are covariantly (as well as gauge-invariantly
and locally) characterized as the case when all perturbed
quantities are expressible as spatial gradients of scalars. In
particular, the nonlocal perturbed bulk effects are described by
$\D_\mu{\cal U}$ and~\cite{m}
\begin{equation}\label{scal}
{\cal Q}_\mu=\D_\mu{\cal Q}\,,~~{\cal P}_{\mu\nu}=\left[
h_\mu{}^\alpha h_\nu{}^\beta-{\textstyle{1\over3}}h^{\alpha\beta}
h_{\mu\nu}\right] \D_{\alpha} \D_{\beta}{\cal P}\,.
\end{equation}
Note that there is no transverse traceless mode from the bulk,
since the nonlocal traceless mode has nonzero spatial
divergence~\cite{m}:
\[
\D^\nu{\cal P}_{\mu\nu}={\textstyle{2\over3}}\D^2(\D_\mu{\cal
P})\,.
\]
The bulk gravitational field affects scalar perturbations via scalar
Coulomb modes, given by the spatial gradients of the `potentials'
${\cal U}$, ${\cal Q}$ and ${\cal P}$.

For adiabatic matter perturbations the 4-acceleration is
\begin{equation}\label{acc}
A_\mu=-{c_{\rm s}^2\over \rho(1+w)}\,\D_\mu\rho\,.
\end{equation}
The gradients
\begin{equation}\label{s3}
\Delta_\mu={a\over\rho}\D_\mu\rho\,,~~Z_\mu=a\D_\mu\Theta\,,
\end{equation}
describe inhomogeneities in the matter and expansion~\cite{ehb},
and the dimensionless gradients describing inhomogeneity in the
nonlocal quantities are~\cite{m}
\begin{equation}\label{s4}
U_\mu={a\over\rho}\D_\mu{\cal U}\,,~Q_\mu={1\over\rho} \D_\mu
{\cal Q}\,,~P_\mu={1\over a\rho}\D_\mu{\cal P}\,.
\end{equation}

The spatial gradient of the conservation equations (\ref{pc1}),
(\ref{lc1'}) and (\ref{lc2'}), and the generalized Raychaudhuri
equation (\ref{prl}), leads to a system of equations for these
gradient quantities~\cite{m}. The gradients define scalars via
their comoving divergences:
\begin{equation}\label{F}
F\equiv a\D^\mu F_\mu\,,~\mbox{ with }~ F=\Delta, Z, U, Q, P\,,
\end{equation}
where $\Delta$ is a covariant analog of the Bardeen density
perturbation $\epsilon_{\rm m}$ (see~\cite{ehb}). Then the system
of equations governing scalar perturbations on the brane follows
from the gradient system given in~\cite{m} as
\begin{eqnarray}
&&\dot{\Delta} =3wH\Delta-(1+w)Z\,,\label{s5}\\ &&\dot{Z}
=-2HZ-\left({c_{\rm s}^2\over 1+w}\right)
\D^2\Delta-\left({6\rho\over\kappa^2\lambda}\right) U\nonumber\\
&&~~{}-{\textstyle{1\over2}}\kappa^2 \rho\left[1+
(4+3w){\rho\over\lambda}- \left({2c_{\rm s}^2\over
1+w}\right){6{\cal U}\over\kappa^4\lambda\rho}\right]
\Delta\,,\label{s6}\\ &&\dot{U} =(3w-1)HU - \left({4c_{\rm
s}^2\over 1+w}\right)\left({{\cal U}\over\rho}\right) H\Delta
\nonumber\\&&~~{} -\left({4{\cal U}\over3\rho}\right)
Z-a\D^2Q\,,\label{s7}\\ &&\dot{Q}
=(1-3w)HQ-{1\over3a}U-{\textstyle{2\over3}} a\D^2P
\nonumber\\&&~~{} +{1\over6a}\left[ \left({8c_{\rm s}^2\over
1+w}\right){{\cal U}\over\rho}-\kappa^4
\rho(1+w)\right]\Delta\,.\label{s8}
\end{eqnarray}

In general relativity, only the first two equations apply, with
$\lambda^{-1}$ set to zero in Eq. (\ref{s6}). In this case we can
decouple the density perturbations via a second-order equation for
$\Delta$, whose independent solutions are adiabatic growing and
decaying modes. Local bulk effects modify the background dynamics,
while nonlocal bulk effects introduce new fluctuations. This leads
to fundamental changes to the simple general relativity picture.
There is no equation for $\dot{P}$, so that in general, scalar
perturbations on the brane cannot be predicted by brane observers
without additional information from the unobservable bulk. Thus in
general, one must solve also the scalar perturbations in the bulk
in order to determine the perturbation evolution on the brane.

However, there is a crucially important exception to this, arising
from the fact that $P$ only occurs in Eqs.~(\ref{s5})--(\ref{s8})
via the Laplacian term $\D^2P$.
We can use the shear propagation equation on the brane~\cite{m} to
provide an order-of-magnitude comparison of $P$ with $\Delta$:
\[
\dot{\sigma}_{\mu\nu}+2H\sigma_{\mu\nu}+E_{\mu\nu}+\D_{\langle\mu}
A_{\nu\rangle}={3\over\kappa^2\lambda}{\cal P}_{\mu\nu}\,,
\]
where $E_{\mu\nu}$ is the electric Weyl tensor on the brane. This
equation, together with Eqs.~(\ref{scal}) and (\ref{acc}) shows that
\[
{1\over\kappa^2\lambda}|\D^2{\cal P}| \sim {1\over\rho}|\D^2\rho|\,,
\]
and then Eqs.~(\ref{s3})--(\ref{F}) imply
\[
|P|\sim {\kappa^2\lambda\over a^2\rho}|\Delta|\,.
\]
Then
\[
|a\D^2P|\sim{k^2\over a^2H^2}\,{\kappa^4\rho\over a}|\Delta|\,,
\]
on using the high-energy Friedmann equation
($H^2\sim\kappa^2\rho^2/\lambda$). Thus, for $k\ll aH$, i.e. on
large scales, well beyond the Hubble horizon, we can neglect the
$\D^2P$ term in Eq.~(\ref{s8}) relative to the  $\Delta$ term.
Thus {\em on large scales, the
system closes on the brane, and brane observers can predict scalar
perturbations from initial conditions intrinsic to the brane,}
without the need to solve the bulk perturbation equations.
Note that the $\D^2Q$ term in Eq.~(\ref{s7}) may also be neglected
relative to the $U$ term. This follows from the shear constraint
equation~\cite{m}
\[
\D^\nu\sigma_{\mu\nu}-{\textstyle{2\over3}}\D_\mu\Theta
=-{6\over\kappa^2\lambda}{\cal Q}_\mu\,,
\]
which gives, on taking the divergence,
\[
|Q|\sim{\kappa^2\lambda\over a\rho}|Z|\sim
{1\over aH}|U|\,,
\]
where the last relation follows from Eq.~(\ref{s6}). The
system Eqs.~(\ref{s5})--(\ref{s8})
then reduces to 3 coupled equations in $\Delta$, $Z$ and
$U$, plus a decoupled equation for $Q$, which determines $Q$ once
the other 3 quantities are solved for. Thus there are in general 3
modes of large-scale density perturbations: {\em a non-adiabatic
mode is introduced by bulk effects.} This mode is carried by
fluctuations $U$ in the nonlocal energy density ${\cal U}$, which
are present even if ${\cal U}$ vanishes in the background. The
fluctuations $Q$ and $P$ in the nonlocal energy flux and
anisotropic stress do not affect the density perturbations on very
large scales.

In qualitative terms, we can interpret these general results as
follows. Bulk effects lead to an effective total energy-momentum
tensor that is non-adiabatic. From Eqs.~(\ref{a}) and (\ref{b}),
we find a measure of the total effective non-adiabatic pressure
perturbation, which is the covariant analog of $\delta p^{\rm
tot}-(c_{\rm s}^2)^{\rm tot}\delta\rho^{\rm tot}$:
\begin{eqnarray}\label{nonad}
&&a\D_\mu p^{\rm tot}-(c_{\rm s}^2)^{\rm tot}a\D_\mu \rho^{\rm
tot}={6H\rho\over \kappa^4\dot{\rho}^{\rm tot}\lambda
}\left(1+{\rho\over\lambda}\right) \times
\nonumber\\&&~{}\times\left[{1\over3}-c_{\rm
s}^2-{\rho+p\over\rho+\lambda}\right] \left\{4{\cal U}\,\Delta_\mu
- 3(\rho+p) U_\mu \right\}\,.
\end{eqnarray}
In addition, Eqs.~(\ref{c}) and (\ref{d}) show that non-adiabatic
stresses and fluxes are also present, due to nonlocal bulk
effects. From this viewpoint, it is not surprising that
non-adiabatic modes arise in the density perturbations.
Furthermore, anisotropic stresses do not affect large-scale
density perturbations,
which can explain how the scalar perturbation equations close on
the brane in this case, since the generalized conservation
equations~(\ref{pc1})--(\ref{lc2'}) form a closed system in the
absence of ${\cal P}_{\mu\nu}$.

Note that an alternative interpretation is also possible. We can
regard the nonlocal bulk effects as constituting a radiative
``Weyl" fluid, with energy-momentum tensor $-\kappa^{-2}{\cal
E}_{\mu\nu}$. This Weyl fluid has non-adiabatic stresses ${\cal
P}_{\mu\nu}$, and is in motion relative to the matter fluid, with
relative velocity parallel to ${\cal Q}_\mu$. Although the matter
fluid obeys energy-momentum conservation, the Weyl fluid does not;
the momentum balance equation~(\ref{lc2'}) shows that density
inhomogeneity in the matter fluid sources a momentum transfer to
the Weyl fluid.

Whatever intuitive interpretation we adopt, the concrete result is
that {\em large-scale density perturbations on the brane may be
determined without knowledge of the bulk, and acquire a
non-adiabatic mode due to effects from the free gravitational
field in the bulk.}

\section{large-scale scalar perturbations}

We rewrite the coupled system for large-scale perturbations by
introducing two useful new quantities. We define,
following~\cite{ehb},
\begin{equation}\label{phi}
\Phi=\kappa^2\rho a^2\Delta\,,
\end{equation}
which is a covariant analog of the Bardeen metric potential
$\Phi_H$, and the covariant local curvature perturbation
\begin{equation}\label{c'}
C=a\D^\mu C_\mu\,, ~C_\mu=a^3\D_\mu{\cal R}\,,
\end{equation}
where ${\cal R}$ is the Ricci curvature of the surfaces orthogonal
to $u^\mu$. (Note that these surfaces are in general shearing, and
non-uniform in $\rho$, $\Theta$, ${\cal U}$ and ${\cal R}$.)

Then the coupled system for density perturbations,
Eqs.~(\ref{s5})--(\ref{s7}), can be rewritten on large scales as
\begin{eqnarray}
\dot{\Phi}&=& -H\left[1+{(1+w)\kappa^2\rho\over 2H^2}\left(1+
{\rho\over \lambda}\right)\right]\Phi \nonumber\\
&&~~{}+\left[{(1+w)\kappa^2 \rho\over 4H}\right]C -
\left[{3(1+w)a^2\rho^2\over \lambda H}\right]U\,, \label{p1}\\
\dot{C} &=& -\left[{72c_{\rm s}^2 H {\cal U}\over (1+w)\lambda
\kappa^4\rho}\right]\Phi\,, \label{p2}\\ \dot{U} &=&
H\left[3w-1-{4{\cal U}\over \kappa^2\lambda
H^2}\right]U+\left({{\cal U}\over 3a^2H\rho}\right) C \nonumber\\
&&~~{}-{2{\cal U}\over 3a^2 H\rho}\left[1+{\rho\over\lambda} +{6
c_{\rm s}^2H^2\over (1+w)\kappa^2\rho}\right]\Phi\,. \label{p3}
\end{eqnarray}

The general relativistic case is recovered when we set
$\lambda^{-1}$, ${\cal U}$ and $U$ to zero; in this case,
Eq.~(\ref{p3}) falls away, and Eq.~(\ref{p2}) reduces to
\begin{equation}\label{dotc}
C=C_o\,,~~\dot{C}_o=0\,,
\end{equation}
which expresses conservation of the covariant curvature
perturbation along each fundamental world-line. The value of $C_o$
will in general vary from world-line to word-line, so that its
conservation is local, and is {\em not} an indicator of purely
adiabatic perturbations. (In general relativity, $\dot{C}=0$ on
large scales for a flat background even when there are
non-adiabatic perturbations~\cite{ehb}.) Bulk effects destroy the
local conservation of $C$ in general, by Eq.~(\ref{p2}).

However, there is an important special case when local
conservation is regained: when the nonlocal energy density ${\cal
U}$ vanishes in the background. This does not mean that
fluctuations in the nonlocal energy density are zero, i.e., we
still have $U\neq 0$ in general. It can be argued that vanishing
${\cal U}$ in the background is more natural, if one believes that
the bulk background should be conformally flat, and thus strictly
anti-de Sitter. (Quantum effects may nucleate a black hole in the
bulk~\cite{gs}, in which case the Schwarzschild-anti de Sitter
bulk, with ${\cal U}_o$ proportional to the black hole mass, would
be a natural background.) From now on, we will assume a
conformally flat background bulk and a spatially flat brane
background; thus ${\cal U}_o=0=K$ in the background generalized
Friedmann equation~(\ref{f}). Equation~(\ref{nonad}) shows that
the non-adiabatic total pressure perturbation is then proportional
to $(\rho/\lambda)U$, which will be enhanced at high energies and
suppressed at low energies.

When ${\cal U}=0$ in the background, Eq.~(\ref{dotc}) holds, and
Eq.~(\ref{p3}) gives
\begin{equation}\label{u}
U=U_of\,,~\dot{U}_o=0\,,~f=\exp \int_{a_o}^a(3w-1)\,d\ln a\,.
\end{equation}
This shows that $U$ rapidly redshifts away during inflation, so
that non-adiabatic effects from nonlocal bulk influence are small.
By contrast, the modifications to the background dynamics from
local bulk effects are strong during inflation at high energy.

The key equation~(\ref{p1}) becomes
\begin{eqnarray}
&&{d\Phi\over dN}+\left[1+{(1+w)\kappa^2\rho\over 2H^2}\left(1+
{\rho\over \lambda}\right)\right]\Phi =\nonumber\\
&&~{}\left[{(1+w)\kappa^2 \rho\over 4H^2}\right]C_o -
\left[{3(1+w)a_o^2\rho^2\over \lambda H^2}\right]e^{2N}fU_o\,,
\label{p1'}
\end{eqnarray}
where $N=\ln(a/a_o)$ is the number of e-folds. We have thus reduced the
coupled system to one simple inhomogeneous linear equation, which
may be integrated along the fundamental world-lines. Along each
world-line, the constancy of $C_o$ and $U_o$ allows us to track
the change in $\Phi$ as $w$ changes, from inflationary behavior
through to radiation- and matter-domination.

We can perform a qualitative analysis of the evolution of $\Phi$
as follows. For high-energy slow-roll inflation, $w$ and $\rho$
are nearly constant, and $V\approx\rho\gg\lambda$, where
$V(\varphi)$ is the inflaton potential. Then Eq.~(\ref{p1'})
implies
\begin{equation}
\mbox{high-energy inflation:}~~~\Phi \approx
{3\over2}(1+w)C_o\,{\lambda\over \rho}\,. \label{sol1}
\end{equation}
By contrast, the general relativity solution is
\begin{equation}\label{sol1'}
\Phi_{\rm gr}\approx {3\over4}(1+w)C_o\,.
\end{equation}
In general relativity, $\Phi$ remains constant on large scales
during slow-roll inflation, independent of the form of the
inflaton potential. In the brane-world, $\Phi$ is {\em slowly
increasing during high-energy slow-roll inflation,} since
$\Phi\sim \rho^{-1}$ and $\rho$ is slowly decreasing. This
qualitative analysis is confirmed by the numerical integration of
a simple phenomenological model shown in Fig.~1. For more
realistic models, i.e., where $V(\varphi)$ is specified, the
evolution of $\Phi$ may be more complicated than shown in Fig.~1.

During reheating, in periods when $w$ is approximately constant on
average (for example, $w\approx0$ for $V={1\over2}m^2\varphi^2$),
Eqs.~(\ref{pc1}) and (\ref{p1'}) imply
\begin{eqnarray}
&&\mbox{high-energy $w\approx$ constant reheating:}\nonumber\\
&&~~~~~\Phi \approx {3(1+w)\over 2(7+6w) }{\lambda\over
\rho_0}C_o\,e^{3(1+w)N}+\,{\rm const}\,. \label{sol2}
\end{eqnarray}
Thus {\em high-energy $w\approx$ constant reheating on the brane
produces amplification of} $\Phi$, unlike general relativity,
where $\Phi$ remains constant on large scales during $w\approx$
constant reheating:
\begin{equation}\label{sol2'}
\Phi_{\rm gr}\approx {3(1+w)\over 2(5+3w)}C_o\,.
\end{equation}

\begin{figure} \begin{center} \includegraphics{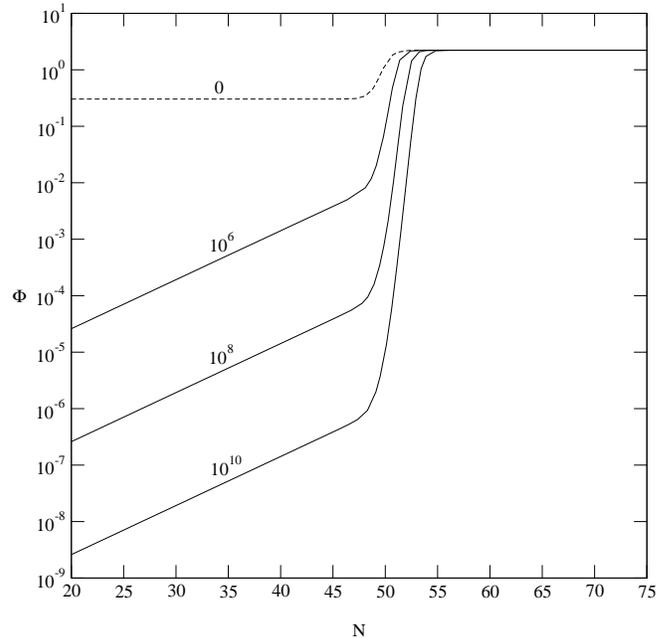} \end{center}
\caption{
The evolution of $\Phi$ along a fundamental world-line for a mode
that is well beyond the Hubble horizon at $N=0$, about 50 e-folds
before inflation ends, and remains super-Hubble through the
radiation era. We have modelled a smooth transition from inflation
to radiation by $w={1\over3}[(2-\alpha)\tanh(N-50)-(1-\alpha)]$,
where $\alpha$ is a small positive parameter (chosen as
$\alpha=0.1$ in the plot). Labels on the curves indicate the value
of $\rho_o/\lambda$, so that the general relativistic solution is
the dashed curve ($\rho_o/\lambda=0$). For $\rho_o/\lambda\gg1$,
Eq.~(\ref{inf}) shows that inflation ends at
$N=50-2\ln[(1-2\alpha)/3]\approx 47.4$, and at $N=50$ in general
relativity. Only the lowest curve still has $\rho/\lambda\gg1$ at
the start of radiation-domination ($N$ greater than about 53), and
one can see that $\Phi$ is still growing, as confirmed by
Eq.~(\ref{sol3}).
}

\end{figure}

In the radiation era, the energy density redshifts rapidly, so
that $\rho$ quickly falls below the brane tension $\lambda$. If
the energy density at the end of reheating is high enough, then at
the start of radiation-domination we have $\rho\gg\lambda$, and we
find that $\Phi$ {\em is amplified during high-energy radiation
domination:}
\begin{equation}
\mbox{high-energy radiation:}~~~\Phi \approx
{2\over9}{\lambda\over \rho_0}\,C_oe^{4N}+\,{\rm const}\,.
\label{sol3}
\end{equation}
At low energies on the brane, or in general relativity, we find
that $\Phi$ is constant:
\begin{equation}\label{sol3'}
\mbox{low-energy radiation:}~~~\Phi\approx\Phi_{\rm gr}\approx
{1\over3}C_o\,.
\end{equation}
This qualitative result is confirmed in Fig.~1. After the
radiation era, the energy scale has fallen well below the brane
tension, so that in the matter era, we recover the general relativity
result:
\begin{equation}\label{sol4}
\mbox{matter era:}~~~\Phi\approx\Phi_{\rm gr}\approx
{3\over10}C_o\,.
\end{equation}

In general relativity, the constancy of $\Phi$ during slow-roll
inflation and radiation- and matter-domination allows one to
estimate the amplification in $\Phi$. CMB large-angle anisotropies
as measured by COBE place limits on the amplified $\Phi$, and this
in turn places constraints on the inflationary potential, since
the potential determines the initial value of $\Phi$. This simple
picture is complicated by high-energy effects in the brane-world.
We have given a rough estimate for the slow-roll inflationary
evolution in Eq.~(\ref{sol1}). In particular, $\Phi$ grows
during inflation, so that placing limits on inflationary
parameters is more complicated.
The evolution of $\Phi$ is also
sensitive to the form of the potential $V(\varphi)$, although
slow-roll conditions will reduce this sensitivity.

\section{Conclusion}

Using the covariant local formalism developed in~\cite{m}, we have
analyzed the evolution of large-scale density perturbations on the
brane. Density inhomogeneity on the brane generates Weyl curvature
in the bulk, which in turn backreacts on the brane, in the form of
a nonlocal energy-momentum tensor. Fluctuations in the nonlocal
energy density induce a non-adiabatic mode in large-scale density
perturbations. The fluctuations in the nonlocal energy flux are
decoupled from the density perturbations, while the nonlocal
anisotropic stress plays no role on large scales. This latter
feature is what closes the system of brane density perturbation
equations, allowing brane observers to evaluate the perturbations
on the brane without solving for the bulk perturbations.

We showed that the local and nonlocal bulk effects arising during
high-energy inflation, and any high-energy start to the radiation
era, modify the simple picture of general relativity. The local
covariant version of the metric perturbation, i.e., $\Phi$, is no
longer constant on large scales during these regimes. Computing
the constraints on inflationary potentials that are imposed by CMB
large-angle anisotropies is therefore more complicated, and more
model-dependent. We gave a rough estimate for slow-roll inflation
in Eq.~(\ref{sol1}); together with Eqs.~(\ref{sol3'}) and
(\ref{sol4}), this indicates how COBE limits can be used to impose
constraints on the inflationary potential. A more accurate
determination of the observational constraints on brane-world
inflation may require numerical integration for the specified form
of $V(\varphi)$, even for large scales. Numerical integration will
also be required for the more complicated case, not investigated
here, when the background bulk is not conformally flat, i.e., when
the nonlocal energy density ${\cal U}$ does not vanish in the
background. In this case, the coupled system of
equations~(\ref{p1})--(\ref{p3}) can no longer be reduced to one
equation, since $C$ and $U$ are no longer locally conserved.

The formalism we have used is restricted to large scales. When a
mode approaches the Hubble radius, the gradient terms can no
longer be neglected, and the presence of these terms means that
the system of equations no longer closes on the brane. A fuller
investigation requires a formalism that can handle all scales, and
which necessarily involves the evolution of perturbations in the
bulk. A covariant formalism for bulk perturbations has not been
developed, but a metric-based formalism has been
developed~\cite{pert,bd}. The equations of this formalism are very
complicated, and considerable work remains to be done before
smaller scale structure can be predicted and compared with
observations of the acoustic peaks in the CMB anisotropies. Our
results provide a useful initial step for further developments by
showing what happens on very large scales.

Although it should be possible to reproduce the Sachs-Wolfe
plateau and satisfy COBE limits by suitable restrictions on
inflationary parameters, it remains to be seen whether this can be
done consistently with the observed small-scale features of CMB
anisotropies and with the observed matter distribution. This is
the basis on which to confront brane-world theories with
cosmological observations.
\[ \]
{\bf Acknowledgements:}\\

We would like to thank Bruce Bassett, Carsten van de Bruck, Gary
Felder and David Wands for helpful discussions and comments.


\end{document}